\documentclass{article}
\usepackage{maa-monthly}

\theoremstyle{theorem}
\newtheorem{theorem}{Theorem}

\theoremstyle{definition}

\newtheorem{lemma}[theorem]{Lemma}

\begin{document}

\title{Derivation of the General Solution of the Black-Scholes Boundary-Value Problem}
\markright{Black-Scholes Boundary Value Problem}
\author{ByoungSeon Choi and M.Y. Choi}

\maketitle

\begin{abstract}
There are infinitely many functions that satisfy
the Black-Scholes partial differential equation
and the terminal condition corresponding to the European call option.
This means that the Black-Scholes formula, which led to the award of the 1997 Nobel Prize in Economic Sciences, is not the unique solution, as was once assumed.
Consequently, it violates the law of one price, one of the fundamental laws of economics and finance.
In this article, we present a rigorous derivation of these solutions to the Black-Scholes boundary value problem.
\end{abstract}

\section{Introduction}
\label{section:Introduction}

In their seminal work, Black and Scholes \cite{BlackScholes1973} derived
the celebrated Black-Scholes formula for the value of a call option in terms of the stock price
under the assumption of ``ideal conditions'' in the stock and option markets:

\begin{description}
\item[a)] The short-term interest rate is known and constant over time.
          The constant interest rate is denoted by $r$.

\item[b)] The stock price follows a random walk in continuous time, and the variance rate is proportional to the square of the stock price.
The variance of the return on the stock is constant. The constant variance is denoted by $v^2$.

\item[c)] The stock pays neither dividends nor other distributions.

\item[d)] The option is ``European'', meaning it can only be exercised at maturity.
The maturity date, also called the expiration date, is denoted by $T$.

\item[e)] There are no transaction costs when buying or selling the stock or option.

\item[f)] It is possible to borrow any fraction of the price of a security at the short-term interest rate in order to buy or hold it.

\item[g)] There are no penalties for short selling.
A seller who does not own a security will simply accept the price of the security from a buyer
and agree to settle with the buyer on a future date by paying her/him an amount equal
to the price of the security on that date.
\end{description}

Under these assumptions, Black and Scholes considered a European call option that pays
$ \left[x_T - K\right]^+ \equiv \max \left\{ x_T{-}K, 0 \right\} $
at expiration date $T$, where $x_{t}$ is the stock price at time $t$,
and the striking price $K$ is a positive constant.
They showed that its fair value $ w( x_t, t)$ satisfies the partial differential equation (PDE):
\begin{equation}
\label{Eq:BS PDE101}
w_{t} ( x , t )= r w ( x , t ) -  r x w_{x} ( x , t )- \frac{1}{2} v^2 x^2 w_{xx} ( x , t ) , \ \
(0 \leq t < T , \ x > 0).
\end{equation}
This is equation 7 from their original paper and is now called the Black-Scholes PDE.
According to the definition of the European call option, its price satisfies the terminal condition:
\begin{equation}
\label{Eq:Terminal Condition101}
\lim\limits_{t \uparrow T} w (x ,t) = \left[ x - K \right]^+, \ \ ( x \neq K).
\end{equation}
Black and Scholes assert that there is only one formula $ w(x_t ,t) $ that satisfies the differential equation
(\ref{Eq:BS PDE101})  subject to the boundary condition (\ref{Eq:Terminal Condition101}),
and they show that the solution is
\begin{equation}
\label{Eq:BS Formula}
w^{BS} (x_t  ,t) = x_t   N ( d_{1} )  - K e^{-r \tau} N (d_2 )
\end{equation}
with $ \tau \doteq T - t $,
where $N( \cdot )$ is the cumulative distribution function of the standard normal random variable with
\begin{align}
 & d_{1} \doteq \frac{1}{v \sqrt{\tau}} \left[ \ln \frac{x_t }{K} + \left( r + \frac{v^2}{2} \right) \tau \right],
   \label{Eq:BS Equation101}  \\
 & d_2 \doteq \frac{1}{v \sqrt{\tau}} \left[\ln \frac{x_t }{K} + \left( r - \frac{v^2}{2} \right) \tau \right] .
   \label{Eq:BS Equation102}
\end{align}
Equation (\ref{Eq:BS Formula}) is the Black-Scholes(-Merton) formula for the price of a European call option, which is perhaps the world's most well-known options pricing model.
Davis and Etheridge \cite[p. 108]{BachelierDavisEtheridge2006} have asserted that the Black and Scholes paper is the ``decisive breakthrough'' that divides the history of option pricing and financial economics into ``the pre-Black–Scholes and post-Black–Scholes eras''.

Contrary to the Black-Scholes formula,
Choi and Choi \cite{ChoiChoi2018} show that
for any $ M \in \mathbb{Z}_{\geq 0} $ and
any $\pmb{\zeta} = \left( \zeta_{0}, \zeta_{2},  \cdots , \zeta_M \right) \in \mathbb{R}^{M+1}$,
\begin{equation}
\label{Eq:CC Formula}
w^{\pmb{\zeta}} (x, t) \doteq  w^{BS} (x,t) + \sum\limits_{m=0}^{M} \zeta_{m} c_{m} (x,t)
\end{equation}
satisfies the Black-Scholes PDE given by Eq. (\ref{Eq:BS PDE101}) together with the terminal condition
in Eq. (\ref{Eq:Terminal Condition101}).
The function $ c_{m} (x,t)  $ will be defined in Section
\ref{section:Deriving the General Solution}.
Clearly, $ w^{\pmb{\zeta}} (x, t)  $ is a general solution
of the Black-Scholes boundary value problem consisting of
the Black-Scholes PDE (\ref{Eq:BS PDE101}) and the terminal condition (\ref{Eq:Terminal Condition101}).
In this article, we provide a thorough solution of the boundary value problem to demonstrate how the formula (\ref{Eq:CC Formula}) is obtained.

\section{Deriving the General Solution}
\label{section:Deriving the General Solution}

In this section, we derive Eq. (\ref{Eq:CC Formula}), which is the general solution
to the boundary value problem consisting
of the Black-Scholes PDE (\ref{Eq:BS PDE101}) and the terminal condition (\ref{Eq:Terminal Condition101}).

It is convenient to change variables as follows:
\begin{align}
 & h_t \doteq \ln \frac{x_t}{K} ,  \label{Eq:NewGS101} \\
 & y(h_t, \tau ) \doteq \frac{1}{K}  w ( x_t, t )
   \exp \left( \frac{k-1}{2} h_t + \frac{ [k + 1 ]^2}{8}v^2 \tau \right)                \label{Eq:NewGS102}
\end{align}
with $k \doteq 2 r / v^2$.
Then, the Black-Scholes PDE (\ref{Eq:BS PDE101}) reduces to the Kolmogorov-Fokker-Planck equation:
\begin{equation}
\label{Eq:Heat PDE102}
y_{\tau} (h_t ,\tau) = \frac{1}{2} v^2 y_{hh} (h_t ,\tau) ,
\end{equation}
which is also known as the heat-transfer equation.
Using the same change of variables, we can express
the terminal condition (\ref{Eq:Terminal Condition101}) as an initial condition
\begin{equation}
\label{Eq:Terminal Condition102}
y(h_T, 0) = \left[\exp \left(\frac{k+1}{2} h_T \right)
- \exp \left( \frac{k-1}{2}h_T \right) \right]^+ .
\end{equation}

Since Fourier \cite{Fourier1822}, it has been known
that the unique solution
of the boundary problem consisting of the Kolmogorov-Fokker-Planck equation (\ref{Eq:Heat PDE102})
and the initial condition  (\ref{Eq:Terminal Condition102}) can be written as
\begin{equation}
\label{Eq:NewGS103}
y (h_t ,\tau) = \int_{-\infty }^{\infty } y(u,0) \, \phi \left(u ; h_t  , v^2 \tau \right) d u ,
\end{equation}
where $\phi (x; \mu, \sigma^2 )$ is the probability density function of the normal
random variable $x$ with mean $\mu$ and variance $ \sigma^2$.
See, \textit{e.g.}, \cite{Evans1998}.
We call $ \phi \left(u ; h_t  , v^2 \tau \right) $ Green's function for this problem.
Black and Scholes \cite{BlackScholes1973} derived the formula  (\ref{Eq:BS Formula})
for the price of a European call option by performing the integration in Eq. (\ref{Eq:NewGS103}).

We now consider the function
\begin{equation}
\label{Eq:NewGS201}
y^{\pmb{\eta}}  (h_t ,\tau) \doteq \int_{-\infty}^{\infty} y(u,0) \,
q^{\pmb{\eta}}  \left(u ; h_t , v^2 \tau \right) d u ,
\end{equation}
where $ q^{\pmb{\eta}}  \left( u ; h_t  ,    v^2 \tau  \right) $ is defined by
\begin{align}
\label{Eq:NewGS202}
q^{\pmb{\eta}}  \left( u ; h_t  ,    v^2 \tau  \right)
\doteq \,  \phi \left( u ; h_t  , v^2 \tau \right)  +
\sum\limits_{l=1}^{M+1} \eta_{l}
 \frac{1 }{( v \sqrt{\tau} )^{l}} He_{l}  \left( \frac{u - h_t  }{v \sqrt{\tau}} \right)
 \phi \left(u ; h_t  , v^2 \tau \right)
\end{align}
with  $ \pmb{\eta} \doteq \left( \eta_{1} ,  \eta_{2} , \cdots , \eta_{M} , \eta_{M+1}\right) \in \mathbb{R}^{M+1} $
and the probabilists' Hermite polynomial
\begin{align}	
\label{GS102}
He_{j} (z) \doteq (-1)^j e^{\frac{z^2}{2}} \frac{d^j}{dz^j} e^{-\frac{z^2}{2}}
=\left(z-\frac{d}{dz} \right)^j \cdot 1     , \ \  \left( j \in \mathbb Z_{\geq 0} \right).
\end{align}
Since $ \int_{-\infty}^{\infty} \exp \left( -\frac{z^2}{2} \right)  He_{l} (z) dz = 0
$ for $ l \in \mathbb Z_{> 0}$,  we know that
\begin{align}
 & \int_{-\infty}^{\infty} q^{\pmb{\eta}}  \left( u ; h_t  ,    v^2 \tau  \right) d u  \nonumber \\
 & = \int_{-\infty}^{\infty}  \phi \left(u ; h_t  , v^2 \tau \right) du +
\sum\limits_{l=1}^{M+1} \eta_{l}
\int_{-\infty}^{\infty}  \frac{1 }{( v \sqrt{\tau} )^{l}} He_{l}  \left( \frac{u - h_t  }{v \sqrt{\tau}} \right)
 \phi \left(u ; h_t  , v^2 \tau \right) du \nonumber \\
 & = 1 .   \label{Eq:NewGS204}
\end{align}
In Appendix A, it is shown that
$ y^{\pmb{\eta}}  (h_t ,\tau) $, which is a generalized function of $ y (h_t ,\tau) $, satisfies the Kolmogorov-Fokker-Planck equation:
\begin{equation}
\label{Eq:Heat PDE102Eta}
y_{\tau}^{\pmb{\eta}}  (h_t ,\tau) = \frac{1}{2} v^2 \, y_{hh}^{\pmb{\eta}}  (h_t ,\tau) .
\end{equation}

We now return to Eq. (\ref{Eq:NewGS201}), which can be decomposed into
\begin{equation}
\label{Eq:NewGS301}
y^{\pmb{\eta}}  (h_t ,\tau) = I_{0} + \sum_{l=1}^{M+1} \eta_{l} I_{l}
\end{equation}
with
\begin{align}
 & I_{l} \doteq \int_{0}^{\infty }  y(u,0) \,
    \frac{1 }{( v \sqrt{\tau} )^{l}} He_{l}  \left( \frac{u - h_t  }{v \sqrt{\tau}} \right)
    \phi \left( u \, ; \, h_{t} , v^{2} \tau \right) \, du , \ \
    	 \left( l \in \mathbb{Z}_{\geq 0 } \right).
                 \label{Eq:NewGS302}
\end{align}
For each $ l $, $ I_{l} $ is a solution
to the Kolmogorov-Fokker-Planck equation (\ref{Eq:Heat PDE102Eta}).
The explicit expression for $I_{l}$ is given in Appendix B.

For each  $ l  \in \mathbb{Z}_{\geq 0 } $, we let
\begin{align}
 J_{l} \doteq  K\exp \left( -\frac{k-1}{2}h_t - \frac{[k+1]^2}{8}v^2 \tau \right)  I_{l} ,
	\label{Eq:NewGS310}
\end{align}
which is a solution of the Black-Scholes PDE (\ref{Eq:BS PDE101}) as implied by Eq. (\ref{Eq:NewGS102}).
It is shown in Appendix C that for each $ l  \in \mathbb{Z}_{> 0 }  $,
\begin{align}
J_{l}  =  & \ \left(  \frac{r}{v^{2}} + \frac{1}{2}  \right)^{l}  J_{0}
     + \left[  \left( \frac{r}{v^{2}} + \frac{1}{2}  \right)^{l}
     - \left( \frac{r}{v^{2}} - \frac{1}{2} \right)^{l}
     		\right]  K e^{-r \tau}  N ( d_{2} )		           \nonumber \\		
     	&	 \ \  +  B_{l} (x_{t} , t ) \,  K e^{-r \tau}  n ( d_{2} )
\label{Eq:NewGS311}  	
\end{align}
with $ n (x) \doteq \phi (x ; 0 ,1) $ and
\begin{align}
B_{l} (x_{t} , t )  \doteq  \sum\limits_{j=1}^{l}
     & \binom{l}{j} \frac{1}{ \left( v \sqrt{\tau} \right)^{j} }	\nonumber \\
   & \times \left[  \left( \frac{r}{v^{2}} + \frac{1}{2} \right)^{l-j} He_{j-1} (-d_{1})
         -  \left( \frac{r}{v^{2}} - \frac{1}{2} \right)^{l-j}   He_{j-1} (-d_{2})  \right] .
\label{Eq:NewGS312}  	
\end{align}
The first five $J_{l}$'s are given by
\begin{align}
 & J_{0} =  \ w^{BS} (x_t , t ) ,                 \label{Eq:NewGS320}  \\
 & J_{1} =  \  \frac{1}{v^2} \left[ r + \frac{1}{2} v^2 \right] J_{0} \;
                   + K e^{-r \tau} N ( d_2 )  , \label{Eq:NewGS321}  \\
 & J_{2} = \ \frac{1}{v^4}  \left[ r + \frac{1}{2} v^2 \right]^{2} J_{0}
   	+ \frac{2 r}{v^2} K e^{-r \tau}  N (d_{2} )
   	+  \frac{1}{v \sqrt{\tau}} K  e^{-r \tau}  n (d_{2} ) , \label{Eq:NewGS322}  \\
 & J_{3} =  \ \frac{1}{v^6}  \left[ r + \frac{1}{2} v^2 \right]^3 J_0 \;
           + \left[ \frac{3 r^2}{v^4} + \frac{1}{4} \right]   K e^{-r \tau} N (d_2 )  \nonumber \\	
  &  \ \ \ \  \ \ \   \ \ \      +  \frac{1}{v^3 \tau \sqrt{\tau}}
  \left[ 2 r \tau - \ln \frac{x_{t}}{K} \right] K  e^{-r \tau} n (d_2) ,
          									 \label{Eq:NewGS323}  \\
 &  J_{4} = \ \frac{1}{v^8}  \left[ r + \frac{1}{2} v^2 \right]^4 J_0 \;
           + \frac{r}{v^{6}}  \left[4 r^2 + v^4  \right]   K e^{-r \tau} N (d_2 )  \nonumber \\
 & \ \ \ \  \ \ \  +  \frac{1}{4 v^5 \tau^{2} \sqrt{\tau}}
   \left\{ \left[ 12 r^2 + v^4 \right] \tau^2 - 4 \left[  v^2 +  2 r \ln \frac{x_{t}}{K} \right] \tau
    + 4 \left[  \ln \frac{x_{t}}{K} \right]^{2}    \right\}  K e^{-r \tau} n (d_2) .  \label{Eq:NewGS324}
\end{align}
Since $ J_{0} $,  $  K e^{-r \tau}  N ( d_{2} )	 $, and $ J_{l} $ for $ l > 0 $ are solutions of
the Black-Scholes PDE (\ref{Eq:BS PDE101}),
Eq. (\ref{Eq:NewGS311}) implies that
$ B_{l} (x_{t} , t ) \,  K e^{-r \tau}  n ( d_{2} ) $  is also a solution.
In summary,
\begin{align}
 w^{\pmb{\eta}}  (x_t , t )  \doteq  w^{BS} (x , t )
  + \sum_{l=1}^{M+1} \eta_{l} B_{l} (x_{t} , t ) \,  K e^{-r \tau}  n ( d_{2} )
	\label{Eq:NewGS331}
\end{align}
is a general solution of the Black-Scholes PDE (\ref{Eq:BS PDE101})
for  any $M $ and any $\pmb{\eta} $.

As shown in Appendix D, we can rewrite $ B_{l} (x_{t} , t ) $ as
\begin{align}
B_{l} (x_{t} , t )
= \sum\limits_{i=0}^{l-1} b_{l,i}
        \frac{1}{ \left( v \sqrt{\tau} \right)^{i+1} }
        He_{i} \left( \frac{1}{v \sqrt{\tau} } \ln \frac{x_{t} }{K}  \right) ,  \label{Eq:CanonicalForm101}
\end{align}
where
\begin{align}
b_{l,i} \doteq \sum\limits_{j=i+1}^{l}  \binom{l}{j} \binom{j-1}{i}  (-1)^{j-1}
       \left[  \left( \frac{r}{v^{2}} + \frac{1}{2} \right)^{l-1-i}
       - \left( \frac{r}{v^{2}} - \frac{1}{2} \right)^{l-1-i}  \right].
  \label{Eq:CanonicalForm102}
\end{align}
Putting Eq. (\ref{Eq:CanonicalForm101}) into Eq. (\ref{Eq:NewGS331}), we can write
the solution in the form
\begin{align}
w^{\pmb{\zeta}}  (x_t , t )
 \doteq   w^{BS} (x_{t} , t ) +  \sum_{m=0}^{M} \zeta_{m}    c_{m} ( x_{t} , t )
 	\label{Eq:CanonicalForm100}
\end{align}
with
\begin{align}
c_{m} ( x_{t} , t )  \doteq
         \frac{1}{ \left( v \sqrt{\tau} \right)^{m+1} }
        He_{m} \left( \frac{1}{v \sqrt{\tau} } \ln \frac{x_{t} }{K}  \right)  e^{-r \tau}   n \left( d_{2} \right)  .
 	\label{Eq:CanonicalForm103}
\end{align}

As shown in Appendix E, for each $ m \in \mathbb{Z}_{\geq 0} $, we have
\begin{align}
 \lim_{\tau \downarrow 0}  c_{m} ( x_{t} , t ) = 0 , \  \ ( x_{t} \neq K ),
	\label{Eq:Terminal104}
\end{align}
and thus
\begin{align}
\lim_{ \tau \downarrow 0} w^{\pmb{\eta}} (x_t , t )
= \lim_{ \tau \downarrow 0}  w^{BS} (x_t , t )
= \left[ x_t - K \right]^{+} , \ \ ( x_{t} \neq K ).
	\label{Eq:Terminal105}
\end{align}
This means that $  w^{\pmb{\eta}}  (x_{t} , t ) $ satisfies the terminal condition (\ref{Eq:Terminal Condition101})
for  any $M $ and any $\pmb{\eta}$.
If $m$ is an odd number, then $  c_{m} ( K , t ) = 0  $.
Thus, if $ \zeta_{0} = \zeta_{2} = \zeta_{4} = \cdots = 0 $,
then the limit in Eq. (\ref{Eq:Terminal105})   holds
even for $ x_{t} = K$. Accordingly,
we have the full solution $w^{\pmb{\zeta}}(x_{t} ,t)=w^{BS} (x_{t} ,t) + \sum\limits_{l=1}^{L} \zeta_{2l-1} c_{2l-1} (x_{t} ,t)$, which is continuous like the Black-Scholes option valuation formula. On the other hand, for $k = 0$, Eq. (\ref{Eq:Terminal104}) does not hold for $x_{t}  = K$, where $c_0 (x_{t} ,t)$ becomes arbitrarily large as $t$ grows toward $T$ (from below). In consequence, $c_0 (x_{t} ,t)$ has discontinuity at $x_{t}  = K$, which reflects the singularity in the terminal condition given by Eq. (\ref{Eq:Terminal Condition101}). Specifically, the term $[x_{t} -K]^+$ in Eq. (\ref{Eq:Terminal Condition101}) is not differentiable at $x_{t}  = K$.

\section{Conclusions and Comments}
\label{section: Conclusions and Comments}

In this paper, we have derived the general solution to the Black-Scholes
boundary value problem for a European option price. This solution includes the Black-Scholes formula as a special case.
This implies that the Black-Scholes formula violates the law of one price (LOOP).

The put-call parity then allows us to obtain the corresponding European put option price $ p^{\pmb{\zeta}}  (x_t , t )  $ as
\begin{align}
 p^{\pmb{\zeta}}  (x_t , t ) & = \ w^{\pmb{\zeta}}  (x_t , t )  - x_t + Ke^{-r \tau} \nonumber \\
   &  = \ w^{BS} (x_t , t )  - x_t +  K e^{-r \tau} +  \sum_{m=0}^{M} \zeta_{m}    c_{m} ( x_{t} , t ) ,
	\label{Eq:Camment101}
\end{align}
where the second equality holds by Eq. (\ref{Eq:CanonicalForm100}). We therefore have
\begin{align}
p^{\pmb{\zeta}}  (x_t , t ) = p^{BS} (x_t , t ) +    \sum_{m=0}^{M} \zeta_{m}    c_{m} ( x_{t} , t )
\label{Eq:Camment102}
\end{align}
for  any $M \in \mathbb Z $ and any $\pmb{\zeta}$,
where $p^{BS} (x_t , t)$ is the Black-Scholes put option price.

\appendix

\renewcommand{\thesection}{\Alph{section}}

\begin{center}
\LARGE\textbf{Appendix}
\end{center}

\section{Proof of Eq. (\ref{Eq:Heat PDE102Eta}).}
\setcounter{equation}{0}
\renewcommand{\theequation}{A.\arabic{equation}}

We can easily show that
$ q^{\pmb{\eta}}   \left( \cdot \, ; \, h_t , v^2 \tau \right) $
satisfies the Kolmogorov-Fokker-Planck equation
\begin{equation}
\label{Eq:CCsolution101}
\frac{\partial q^{\pmb{\eta}}   \left( u \, ; \, h_t , v^2 \tau \right) }{\partial \tau}
= \frac{1}{2} v^2 \frac{\partial^2 q^{\pmb{\eta}}   \left( u\, ; \, h_t , v^2 \tau \right) }{\partial h_{t}^{2} } .
\end{equation}
Defining 
\begin{equation}
\label{Eq:CCsolution102}
z^{\pmb{\eta}} (h_t ,\tau)
\doteq \int_{-\infty}^{\infty} y(u,0) \, q^{\pmb{\eta}} \left(u ; h_t , v^2 \tau \right) d u ,
\end{equation}
we know that
\begin{align}
 & \ \frac{\partial z^{\pmb{\eta}} (h_t ,\tau) }{\partial \tau}
 =  \int_{-\infty}^{\infty} y(u,0) \frac{\partial q^{\pmb{\eta}} \left(u ; h_t , v^2 \tau \right)
 													}{\partial \tau} d u \nonumber \\
 = & \  \int_{-\infty}^{\infty} y(u,0) \frac{1}{2} v^2 \frac{\partial^2 q^{\pmb{\eta}}   \left( u\, ; \, h_t , v^2 \tau \right) }{\partial h_{t}^{2} } d u \nonumber \\
 = & \  \frac{1}{2} v^2 \frac{\partial^2}{\partial h_{t}^{2} } \int_{-\infty}^{\infty} y(u,0)
     q^{\pmb{\eta}}   \left( u\, ; \, h_t , v^2 \tau \right)  d u ,
      \label{Eq:CCsolution103}
\end{align}
where the first equality holds by Eq. (\ref{Eq:CCsolution102})
and the second one by Eq. (\ref{Eq:CCsolution101}).
We thus have
\begin{align}
 \frac{\partial z^{\pmb{\eta}} (h_t ,\tau) }{\partial \tau}
 = & \  \frac{1}{2} v^2 \frac{\partial^2 z^{\pmb{\eta}} (h_t ,\tau) }{\partial h_{t}^{2} }  ,
       \label{Eq:CCsolution104}
\end{align}
which manifests that $ y^{\pmb{\eta}} (h_t ,\tau) $ in Eq. (\ref{Eq:NewGS201})
satisfies  the Kolmogorov-Fokker-Planck equation (\ref{Eq:Heat PDE102Eta}).

\section{Explicit expression of $I_{\MakeLowercase{l}}$.}
\setcounter{equation}{0}
\renewcommand{\theequation}{B.\arabic{equation}}

We need some well-known properties of the Hermite polynomials
\cite[Chapter 22]{AbramowitzStegun1972}.

\begin{lemma}
\label{lemma:HermitePolynomials101}
For each $l \in \mathbb Z_{> 0} $, the following holds;
\begin{description}
\item[a)]  $ He_{l+1} (x) = x He_{l} (x) -  l He_{l-1}(x) , $

\item[b)]  $ He_{l} (x) n (x) = (-1)^{l} \frac{d^{l} n (x)}{d x^{l} } ,  $

\item[c)]  $ \int_{d}^{\infty} He_{l} (x) n (x) dx = He_{l-1} (d) n (d) ,  $

\item[d)]  $ He_{l} (x+a) = \sum\limits_{j=0}^{l} \binom{l}{j} a^{l-j} He_{j} (x) $.
\end{description}
\end{lemma}
\vspace{0.3cm}

Putting Eq. (\ref{Eq:Terminal Condition102}) into Eq. (\ref{Eq:NewGS103}) yields
\begin{equation}
\label{Eq:AppenA321}
I_{l} = I_{l,1} - I_{l,2} ,
\end{equation}
where
\begin{align}
 & I_{l,1} \doteq \int_{0}^{\infty }    \exp \left( \frac{k+1}{2} u \right)
    \frac{1 }{( v \sqrt{\tau} )^{l}} He_{l}  \left( \frac{u - h_t  }{v \sqrt{\tau}} \right)
    \phi \left( u \, ; \, h_{t} , v^{2} \tau \right) \, du ,
 	\label{Eq:AppenA322}  \\
 & I_{l,2} \doteq \int_{0}^{\infty }    \exp \left( \frac{k-1}{2} u \right)
    \frac{1 }{( v \sqrt{\tau} )^{l}} He_{l}  \left( \frac{u - h_t  }{v \sqrt{\tau}} \right)
    \phi \left( u \, ; \, h_{t} , v^{2} \tau \right) \, du .
 	\label{Eq:AppenA323}
\end{align}
The change of variable $z \doteq [ u - h_t  ]/[ v \sqrt{\tau} \,] $ in Eq. (\ref{Eq:AppenA322}) yields
\begin{align} 	
 I_{l,1} =  \frac{1}{ ( v \sqrt{\tau} )^{l} } \exp \left( \frac{k+1}{2}h_t \right)
            \int_{-h_t /[ v \sqrt{\tau} \,] }^{\infty } He_{l} (z)
            \frac{1}{\sqrt{2 \pi }} \exp \left( \frac{k+1}{2}v \sqrt{\tau} z - \frac{z^2}{2}\right) \, dz ,
 	\label{Eq:AppenA324}
\end{align}
which, upon further change of variable $x \doteq z - v \sqrt{\tau} [ k+1 ]/2 $, becomes
\begin{align}
 I_{l,1} = &  \frac{1}{( v \sqrt{\tau} )^{l}} \exp \left( \frac{k+1}{2} h_t   + \frac{[ k+1 ]^2}{8} v^2 \tau \right)
  \int_{- d_1 }^{\infty } He_{l} \left( x+ \frac{k+1}{2} v \sqrt{\tau} \right) n (x)  \, dx .
 	\label{Eq:AppenA325}
\end{align} 	
Applying Lemma \ref{lemma:HermitePolynomials101}(d) to Eq. (\ref{Eq:AppenA325}), we obtain
\begin{align}
 I_{l,1} = &  \frac{1}{( v \sqrt{\tau} )^{l}} \exp \left( \frac{k+1}{2} h_t   + \frac{[ k+1 ]^2}{8} v^2 \tau \right)
 										\nonumber \\
     & \times    \sum\limits_{j=0}^{l} \binom{l}{j} \left( \frac{k+1}{2} v \sqrt{\tau} \right)^{l-j}
   		\int_{- d_1 }^{\infty } He_{j} (x)  n (x)  \, dx .
 	\label{Eq:AppenA326}
\end{align} 	
Applying Lemma \ref{lemma:HermitePolynomials101}(c) to Eq. (\ref{Eq:AppenA326}) then leads to
\begin{align}
 I_{l,1} = &  \frac{1}{ ( v \sqrt{\tau} )^{l} }
 		\exp \left( \frac{k+1}{2} h_t   + \frac{[ k+1 ]^2}{8} v^2 \tau \right)             \nonumber \\
 & \ \times \left[ \left( \frac{k+1}{2} v \sqrt{\tau} \right)^{l}  N ( d_{1} )
 +      \sum\limits_{j=1}^{l} \binom{l}{j} \left( \frac{k+1}{2} v \sqrt{\tau} \right)^{l-j}
   		 He_{j-1} (-d_{1})  n (d_{1})  \right] .
 	\label{Eq:AppenA327}
\end{align}

Similarly, it is straightforward to show that Eq. (\ref{Eq:AppenA323}) reduces to
\begin{align} 	
 I_{l,2} =  & \frac{1}{( v \sqrt{\tau} )^{l}} \exp \left( \frac{k-1}{2}h_t \right) \nonumber \\
            & \times \int_{-h_t /[ v \sqrt{\tau} \,] }^{\infty } He_{l} (z)
            \frac{1}{\sqrt{2 \pi }} \exp \left( \frac{k-1}{2}v \sqrt{\tau} z - \frac{z^2}{2}\right) \, dz ,
 	\label{Eq:AppenA334}
\end{align}
which reads
\begin{align}
 I_{l,2} = &  \frac{1}{( v \sqrt{\tau} )^{l}}
              \exp \left( \frac{k-1}{2} h_t   + \frac{[ k-1 ]^2}{8} v^2 \tau \right)     \nonumber \\
 & \ \times \left[ \left( \frac{k-1}{2} v \sqrt{\tau} \right)^{l}  N ( d_{2} )
    + \sum\limits_{j=1}^{l} \binom{l}{j} \left( \frac{k-1}{2} v \sqrt{\tau} \right)^{l-j}
   		He_{j-1} (-d_{2})  n (d_{2})  \right] .
 	\label{Eq:AppenA337}
\end{align}

\section{Explicit expression of $J_{\MakeLowercase{l}}$.}
\setcounter{equation}{0}
\renewcommand{\theequation}{C.\arabic{equation}}

We begin with the following lemma, which is easy to prove.

\begin{lemma}
\label{lemma:Canonical101}

The following equations hold:
\begin{description}
\item[a)] $ K \exp \left( -\frac{k-1}{2}h_t - \frac{[ k+1 ]^2}{8} v^2 \tau \right)
\exp \left(\frac{k+1}{2}h_t + \frac{[ k+1 ]^2}{8} v^2 \tau \right)= x_{t} $

\item[b)]  $ K \exp \left( -\frac{k-1}{2}h_t - \frac{[ k+1 ]^2}{8} v^2 \tau \right)
\exp \left(\frac{k-1}{2}h_t + \frac{[ k-1 ]^2}{8} v^2 \tau \right) = K e^{-r \tau}  $

\item[c)] $ x_t n \left( d_1 \right) = K e^{-r \tau} n \left( d_2 \right) $
\end{description}
\end{lemma}

To derive $J_{l} $, we first define
\begin{align}
  J_{l,m} \doteq K \exp \left(-\frac{k-1}{2} h_t - \frac{[ k+1 ]^2}{8} v^2 \tau \right)  I_{l,m} ,
  \ \ \  ( m = 1, 2 ) .
  \label{Eq:AppenB201}
\end{align}
Putting Eqs. (\ref{Eq:AppenA327}) and (\ref{Eq:AppenA337}) into Eq. (\ref{Eq:AppenB201})
and using Lemma \ref{lemma:Canonical101}, we get
\begin{align}
 J_{l,1}  =  & \ x_{t} \frac{1}{ (v \sqrt{\tau})^{l} }
          \left[ \left( \frac{k+1}{2} v \sqrt{\tau} \right)^{l}  N ( d_{1} )  \right. \nonumber \\
    & \ \ \  \ \ \ \ \ \ \ \ \  \ \ \ \ \ \ \  \ \ \ \ \
    +  \left. \sum\limits_{j=1}^{l} \binom{l}{j} \left( \frac{k+1}{2} v \sqrt{\tau} \right)^{l-j}
   		 He_{j-1} (-d_{1})  n (d_{1})  \right]  ,  \label{Eq:AppenB202}  \\
 J_{l,2} = & \ K e^{-r \tau} \frac{1}{ (v \sqrt{\tau})^{l} }
       \left[ \left( \frac{k-1}{2} v \sqrt{\tau} \right)^{l}  N ( d_{2} )  \right. \nonumber \\
   & \ \ \  \ \ \ \ \ \ \ \ \  \ \ \ \ \ \ \  \ \ \ \ \
    +  \left.  \sum\limits_{j=1}^{l} \binom{l}{j} \left( \frac{k-1}{2} v \sqrt{\tau} \right)^{l-j}
   		He_{j-1} (-d_{2})  n (d_{2})  \right] , \label{Eq:AppenB203}   	
\end{align}
which lead to
\begin{align}
 J_{l} = & K\exp \left( -\frac{k-1}{2}h_t - \frac{[k+1]^2}{8}v^2 \tau \right)  I_{l}  \nonumber \\
 = & K \exp \left( -\frac{k-1}{2}h_t - \frac{[k+1]^2}{8}v^2 \tau \right)  (I_{l,1} - I_{l,2} )\nonumber \\
 = & J_{l,1} - J_{l,2}                      \nonumber \\
 = & \ \left(  \frac{r}{v^{2}} + \frac{1}{2}  \right)^{l}
           \left\{ x_{t}  N ( d_{1} ) - K e^{-r \tau}  N ( d_{2} ) \right\}   \nonumber \\
   &   \ \ + \left\{  \left( \frac{r}{v^{2}} + \frac{1}{2}  \right)^{l}
     - \left( \frac{r}{v^{2}} - \frac{1}{2} \right)^{l}
     		\right\}  K e^{-r \tau}  N ( d_{2} )						\nonumber \\
  & \ \ + K e^{-r \tau}  n ( d_{2} )
	\sum\limits_{j=1}^{l}  \binom{l}{j} \frac{1}{ \left( v \sqrt{\tau} \right)^{j} }
      \left[  \left( \frac{r}{v^{2}} + \frac{1}{2} \right)^{l-j} He_{j-1} (-d_{1}) \right. \nonumber \\
  &   \ \ \  \ \ \ \ \ \ \ \ \  \ \ \ \ \ \ \  \ \ \ \ \
  	 \ \ \  \ \ \ \ \ \ \ \ \  \ \ \ \ \ \ \  \ \ \ \ \  \ \ \ \  \ \ \ \ \
  \left.  -  \left( \frac{r}{v^{2}} - \frac{1}{2} \right)^{l-j}   He_{j-1} (-d_{2})  \right] .
 \label{Eq:AppenB204}   	
\end{align}
The first equality holds by Eq. (\ref{Eq:NewGS310}), the second by Eq. (\ref{Eq:AppenA321}),
the third by Eq. (\ref{Eq:AppenB201}), and the last by Eqs. (\ref{Eq:AppenB202})
and (\ref{Eq:AppenB203}).

\section{Canonical Form.}
\setcounter{equation}{0}
\renewcommand{\theequation}{D.\arabic{equation}}

We know that
\begin{align}
 &  He_{k} \left( -d_{1} \right) = (-1)^{k} He_{k} \left( d_{1} \right)      \nonumber \\
 &  =  (-1)^{k} \sum\limits_{i=0}^{k} \binom{k}{i}
       \left[ \left( \frac{r}{v^{2}} + \frac{1}{2} \right) ( v \sqrt{\tau} ) \right]^{k-i}
        He_{i} \left( \frac{1}{v \sqrt{\tau} } \ln \frac{x_{t} }{K}  \right) ,  \label{Eq:Canonical101}
\end{align}
where the first equality holds by the symmetry of the Hermite polynomials, and the second one
by Lemma \ref{lemma:HermitePolynomials101}(d).
Thus, we have
\begin{align}
& \sum\limits_{j=1}^{l}  \binom{l}{j} \frac{1}{ \left( v \sqrt{\tau} \right)^{j} }
       \left( \frac{r}{v^{2}} + \frac{1}{2} \right)^{l-j} He_{j-1} (-d_{1})   \nonumber \\
= & \sum\limits_{j=1}^{l}  \binom{l}{j} \frac{1}{ \left( v \sqrt{\tau} \right)^{j} }
       \left( \frac{r}{v^{2}} + \frac{1}{2} \right)^{l-j}        (-1)^{j-1}   \nonumber \\
  &  \ \ \ \ \  \ \ \ \ \times     \sum\limits_{i=0}^{j-1} \binom{j-1}{i}
       \left[ \left( \frac{r}{v^{2}} + \frac{1}{2} \right) ( v \sqrt{\tau} ) \right]^{j-1-i}
        He_{i} \left( \frac{1}{v \sqrt{\tau} } \ln \frac{x_{t} }{K}  \right) ,  \label{Eq:Canonical102}
\end{align}
which can be written as
\begin{align}
& \sum\limits_{j=1}^{l}  \binom{l}{j} \frac{1}{ \left( v \sqrt{\tau} \right)^{j} }
       \left( \frac{r}{v^{2}} + \frac{1}{2} \right)^{l-j} He_{j-1} (-d_{1})   \nonumber \\
= &  \sum\limits_{i=0}^{l-1} \left[ \sum\limits_{j=i+1}^{l}  \binom{l}{j} \binom{j-1}{i}  (-1)^{j-1}
       \left( \frac{r}{v^{2}} + \frac{1}{2} \right)^{l-1-i}  \right]
        \frac{1}{ \left( v \sqrt{\tau} \right)^{i+1} }
        He_{i} \left( \frac{1}{v \sqrt{\tau} } \ln \frac{x_{t} }{K}  \right) .  \label{Eq:Canonical103}
\end{align}
Using the same method, we can show
\begin{align}
& \sum\limits_{j=1}^{l}  \binom{l}{j} \frac{1}{ \left( v \sqrt{\tau} \right)^{j} }
       \left( \frac{r}{v^{2}} - \frac{1}{2} \right)^{l-j} He_{j-1} (-d_{2})   \nonumber \\
= & \sum\limits_{i=0}^{l-1} \left[ \sum\limits_{j=i+1}^{l}  \binom{l}{j} \binom{j-1}{i}  (-1)^{j-1}
       \left( \frac{r}{v^{2}} - \frac{1}{2} \right)^{l-1-i}  \right]
        \frac{1}{ \left( v \sqrt{\tau} \right)^{i+1} }
        He_{i} \left( \frac{1}{v \sqrt{\tau} } \ln \frac{x_{t} }{K}  \right) .  \label{Eq:Canonical104}
\end{align}
Putting Eqs. (\ref{Eq:Canonical103}) and (\ref{Eq:Canonical104}) into Eq. (\ref{Eq:NewGS312}) yields
\begin{align}
B_{l} (x_{t} , t )
= \sum\limits_{i=0}^{l-1} b_{l,i}
        \frac{1}{ \left( v \sqrt{\tau} \right)^{i+1} }
        He_{i} \left( \frac{1}{v \sqrt{\tau} } \ln \frac{x_{t} }{K}  \right) ,  \label{Eq:Canonical105}
\end{align}
which completes a derivation of Eq. (\ref{Eq:CanonicalForm101}).

\section{Terminal Condition.}
\setcounter{equation}{0}
\renewcommand{\theequation}{E.\arabic{equation}}

We know that, for $ x_t \neq K $,
\begin{align}
 &  \lim_{ \tau \downarrow 0}    e^{-r \tau}  n ( d_{2} )
      \frac{1}{ \left( v \sqrt{\tau} \right)^{l} }
        He_{l-1} \left( \frac{1}{v \sqrt{\tau} } \ln \frac{x_{t} }{K}  \right)\nonumber \\
=  \ &  \lim_{ \tau \downarrow 0}  \frac{1}{ \left( v \sqrt{\tau} \right)^{l} }
\left( \frac{1}{v  \sqrt{\tau}}   \ln \frac{x_{t}}{K}  \right)^{l-1} \frac{1}{ \sqrt{2 \pi} }
 \exp \left( - \frac{1}{2 v^2}  \left[ \ln  \frac{x_t }{K} \right]^2 \frac{1}{\tau} \right)  \nonumber \\
 = \ &  \frac{1}{\sqrt{2 \pi} v^{2l-1} }  \left(   \ln \frac{x_{t}}{K}  \right)^{l-1}
 \lim_{ \tau \downarrow 0}    \frac{1}{\tau^{l-1} \sqrt{\tau}}
              \exp \left( - \frac{1}{2 v^2}  \left[ \ln  \frac{x_t }{K} \right]^2 \frac{1}{\tau} \right)  \nonumber \\
 = \ & \frac{1}{\sqrt{2 \pi} v^{2l-1}}  \left(   \ln \frac{x_{t}}{K}  \right)^{l-1}  \lim_{u \rightarrow \infty}
 \frac{u^{2l-1}}{\exp \left(  \frac{1}{2 v^2}  \left[ \ln  \frac{x_t }{K} \right]^2  u^2 \right) }
 = 0 ,                 	\label{Eq:ZTcondition111}
\end{align}
where the third equality is obtained via the change of variable $ u \doteq 1/\sqrt{\tau} $,
and the last one can be proven through the use of L'Hospital's rule.
This completes a derivation of Eq. (\ref{Eq:Terminal104}).

\begin{biog}
\item[ByoungSeon Choi] received his Ph.D. in Statistics and Economics in 1983 from Stanford University.
He was a professor of Applied Statistics at Yonsei University and
held visiting positions at Stanford University, UC Santa Barbara, UC San Diego, UC Berkeley, and University of Tokyo.
\begin{affil}
Department of Economics, Seoul National University, Seoul 08826, Korea\\
bschoi12@snu.ac.kr
\end{affil}

\item[M.Y. Choi] received his Ph.D. in Applied Physics in 1985 from Stanford University. He has held visiting positions at various institutes including Pohang University of Science and Technology, Korea Institute for Advanced Study, University of Washington, Ohio State University, Carnegie Mellon University, Los Alamos National Laboratory, Centre National de la Recherche Scientifique, and Henri Poincar\'e Universit\'e.
\begin{affil}
Department of Physics and Center for Theoretical Physics, Seoul National University, Seoul 08826, Korea\\
mychoi@snu.ac.kr
\end{affil}
\end{biog}
\vfill\eject

\end{document}